# A genetic algorithm for straight-line embedding of a cycle onto a given set of points inside simple polygons

Maryam Fadavian[1] , Heidar Fadavian[2]

[1] Department of Computer Engineering & IT, Amirkabir University of Technology, Tehran, Iran
[2] Department of Electrical and Computer Engineering, Tarbiat Modares University, Tehran, Iran

**Abstract:** In this paper, we have examined the problem of embedding a cycle of n vertices onto a given set of n points inside a simple polygon. The goal of the problem is that the cycle must be embedded without bends and does not intersect itself and the polygon. This problem is a special case of the open problem of finding a (s, X, t) - simple Hamiltonian path inside a simple polygon that does not intersect itself and the sides of the polygon. The complexity of the problem is examined in this paper is open, but it has been proved that similar problems are NP-complete. We have presented a metaheuristic algorithm based on a genetic algorithm for straight-line embedding of a cycle with the minimum numbers of intersections, onto a given set of points inside simple polygons. The efficiency of the proposed genetic algorithm is due to the definition of the mutation operation, which removes it if there is an intersection between the embedded edges of the cycle. The experimental results show that the results of the version of the algorithm that uses this mutation operation are much more efficient than the version that uses only the usual two-points mutation operation.

## 1  Introduction

The problem of embedding a graph on a set of points is one of the well-known problems in the field of graph theory and computational geometry. In this case, considering a given graph and a given set of points, the vertices of the graph should be mapped onto the points such that some constraints are satisfied and some objectives are optimized. For example, one of the constraints may be the planarity of the embedding, i.e. the edges of the embedded graph should not intersect, or the edges should be straight lines. Another objective that can be optimized is to minimize the sum of the embedded graph edges [1].

This problem has many applications, for example, the well-known problem of the Euclidean traveling salesman (TSP) can be considered as a specific case of the point-set embedding of graphs, such that the sum of Euclidean length of embedded edges is minimized. Numerous studies have been done on different versions of this issue. But only a few of them have considered the polygon constraint enclosing points. If we consider the constraint of the enclosing polygon, if we want the embedding to be planar, then the edges of the embedded graph should not only intersect each other but also should not intersect the sides of the polygon. In this case, it may not be possible to embed a graph on the given points inside the polygon in a straight line, so in some cases, there may be an intersection either the embedding of some edges is not a straight line, i.e. an edge is embedded as a poly-line. The fracture site in the poly-line embedding of an edge is called a "bend". Therefore, in this case, one of the objectives that can be minimized is the number of curves of the edges.

In this paper, the problem of "a straight-line embedding of a cycle on given points inside a simple polygon" is examined. More precisely, the problem is that by having a simple general polygon $Q$ with m vertices and a set $X$ of $n$ points, which are inside the polygon, we want to embed the simple cycle $P$ of length $n$ on the points $X$ such that the vertices of $P$ are on the points $X$ and $P$ does not intersect its edges and the sides of the polygon $Q$, and at

✉ Heidar Fadavian
   E-Mail: heidarfadavian@modares.ac.ir

Maryam Fadavian
E-Mail: m.fadavian@aut.ac.ir

the same time the edges of the cycle are straight (without bending). In this problem, the number of vertices $P$ is equal to the number of points $X$ and is equal to $n$, so all $n$ points of the set $X$ must be used.

One of the applications of this problem is in the field of robotics. In this way, a robot wants to reach a destination from one source to another by passing a certain number of stations. Now we have to find the path of the robot such that it has the minimum rotation and change of direction. Another application of this problem is in the design of electronic circuits. In this case, places for the circuit components have been specified and routing should be done in such a way that the connections have the minimum bending.

In this paper, we have presented a genetic algorithm for solving this problem. The efficiency of the proposed genetic algorithm is due to the definition of the mutation operation, which removes it if there is an intersection between the embedded edges of the cycle. The experimental results show that the results of the version of the algorithm that uses this mutation operation are much more efficient than the version that uses only the usual two-points mutation operation.

In the case of geometric embedding of a graph, a graph with $n$ nodes and a set of $n$ points on a plane is given, and the goal is to find a single map between the vertices of the graph and the set of points. This is one of the most important issues in computational geometry. Many studies in this field have embedded graphs on the points of the plane, but there are few researchers that consider the constraint of the polygon enclosing the points and embed them inside the polygon. There are different versions of graph embedding, each with different goals. In versions that embed on the page, the goal is generally to minimize the length of the embedded graph edges, and in versions that embed the points inside the polygon, the goal is usually to minimize the curvature of the embedded graph edges. In recent years, algorithms for embedding graphs in unlimited space and limited space have been proposed with the goal of minimizing the length and curvature of edges. Also, in the field of embedding the tree on the points inside the polygon, algorithms that embed the tree planar and with a small number of bends are presented. Among the works that have been done in the field of embedding graphs in unlimited space, the following can be mentioned:

- Provision an optimal algorithm for straight-line planar embedding of a tree on a set of points [2].
- Provision of an algorithm for straight-line planar embedding of an outerplanar graph on a set of points [3].
- Provision of an algorithm for determining the presence or absence of a straight-line embedding of a planar 3-tree $G$ with $n$ vertices on the set $S$ of $n$ points, which gives such an embedding if present [4].
- Provision of two heuristic algorithms to solve the problem of embedding a tree on a set of points on the plane with the shortest length, such that the number of points inside the page is equal to the given tree nodes and the embedded tree is Isomorphic [5].
- Provision of a number of heuristic algorithms to embed a tree on a set of points with the minimum total length [1].
- Proof of NP-completeness of the Simultaneous embedding of two simple separate paths where all the vertices of one of the paths are red and all the vertices of the other path are blue, i.e. the red vertices are embedded on the red points and the blue vertices are embedded on the blue points. Such that the paths do not intersect and provide a heuristic algorithm with an average time order of $O(n^3)$ to solve it [6].

Also, among the works that have been done in the field of embedding graphs in limited space, the following can be mentioned:

- Provision of an algorithm for embedding a tree on points inside a simple polygon. This algorithm is an heuristic algorithm and ensures that each edge does not bend more than $r / 2$, where $r$ is the number of non-convex vertices of a polygon [7].
- Provision an algorithm for embedding a tree with $n$ vertices on $n$ points inside a polygon with m vertices such that the number of bends of the resulting tree is minimized [8].
- Provision an algorithm with time complexity $O(log\ m\ (n^2 + m))$ and space $O(n^2 + m)$ to embed an edge on an $X$ set of $n$ points inside a simple polygon $Q$ with $m$ vertices, whose vertices are the beginning and end of the edge Must be mapped at two specific points $s, t \in X$. The embedded edge should either be a straight line or have the minimum number of bends and these bends are located only at points $X$, also the embedded edge should not cross the boundaries of $Q$ and itself [9]. This algorithm improves the previous two algorithms [10], [11], one with time and space $O(m^2 n^2)$ and the other with time $O(n^3 log\ m + mn)$ and space $O(n^3 + m)$.
- Provision a meta-heuristic algorithm for embedding a $(s, X, t)$-path of specified length $k\ (k \leq n)$ on a set of $X$ of $n$ points inside a simple polygon $Q$ with $m$ vertices with the minimum number of bends and intersections Which intersects the boundaries of $Q$, which is an extension of the work done in [9], because in this paper an edge is embedded and the curves of that edge are minimized (because only the two definite points $s$ and $t$ must be in the path to be) [12].

It should be noted that finding a $(s, X, t)$-Hamiltonian path, which is not necessarily a simple path, inside a simple polygon is an NP-complete problem. This proof is also true when the set of $X$ points is bounded by polygons. Note that the polygon boundary is not an answer because $s$ and $t$ are not necessarily consecutive [13]. Also finding a $(s, X, t)$-simple path with a set of arbitrary obstacles (not a simple polygon) such that the path obtained does not collide with any of the obstacles is an NP-complete problem [10]. Finding a $(s, X, t)$-simple Hamiltonian path is an NP-complete problem despite the arbitrary obstacle such that the path obtained does not collide with the obstacle [6].

## 2    Proposed Genetic Algorithm

In this paper, we present a genetic algorithm for straight-line embedding of a cycle with the minimum number of intersections on the points inside simple random polygons. For this purpose, we define two different mutation operations. The first mutation we define is a common two-point mutation that randomly changes two genes on a chromosome. The second mutation operation we define is the operation defined specifically for this graph embedding problem. This operation is defined in such a way that from generation to generation leads to a reduction of intersections in the embedding of the graph. In order to evaluate the performance of these two mutation operations, we proposed two versions of the algorithm for the problem, the first version of the algorithm using only the usual mutation but the second version of the algorithm using both the usual mutation and the second mutation. The results of the experiments show that the second version of the algorithm, which uses the second mutation operation, has been very successful. In the following, we will explain how to model the problem and solve it by the method of genetic algorithm.

### 2.1    Definition of chromosomes and genes

In the proposed method, each embedding of a cycle on the points inside the polygon is equivalent to one chromosome, and each vertex mapping of the cycle to a certain point is equivalent to one gene on this chromosome, which in fact each gene keeps $x$ and $y$ coordinates of the point that each vertex of the cycle is embedded on it. Figure. 1 shows an overview of this chromosome and its genes for a cycle with 6 vertices.

### 2.2    Initial population

In the proposed algorithm, the initial population is generated randomly with a uniform random distribution. Each person in this population is a cycle embedded in the given points inside the polygon. In fact, to produce each person (each embedded cycle) we have points with a number of vertices equal to $n$, now we connect these points in a straight line sequence to return to the beginning vertex. In this way, we produce a straight-line embedding of the cycle. The initial population consists of a certain number of individuals (embedded cycle) Which are produced in this way.

### 2.3    Crossover operation

The crossover operation performed by first randomly selecting two parent chromosomes from the current parent population. The number of genes on each chromosome is equal to n, we generate a random number r between one and $n$ as the position of the combination of the two parents. From two parent chromosomes we produce a child. We select genes 1 to r-1 from the first parent chromosome as genes 1 to r-1 of the child, and genes of the second part of the child chromosome are selected from genes 1 to $n$ of the second parent chromosome, so that the newly selected genes Have not been selected in the first part of the child chromosome [12]. Figure. 2 shows how to perform the crossover operation. To produce children, a certain number of crossover operations are performed.

### 2.4    Mutation operation

In the proposed algorithm, two different mutation operations are used. The first mutation operation used is the two-point mutation operation. To perform this mutation, a parent is first randomly selected from the parent population. Then a gene from the parent chromosome (a mapped vertex of the cycle) is randomly selected and its value is exchanged with another random gene from the parent chromosome. Figure. 3 shows how to perform the mutation operation.

The second mutation is performed as follows. First, a parent is randomly selected from the parent population. This parent is actually an embedding of the given cycle. We examine the edges of this embedding to find two intersecting edges. We define the mutation operation in such a way that this intersection is eliminated. To remove the intersection of these two edges, remove these two edges and add two more edges to the cycle. Consider Figure. 4. In this figure, the two edges $(x_2, x_3)$ and $(x_6, x_7)$ intersect. In this figure, if we remove these two edges from a cycle and add two edges $(x_2, x_6)$ and $(x_3, x_7)$ to the cycle, the intersection of those two edges will be removed. To perform this operation, we first find the two vertices $x_2$ (the beginning edge of the edge $(x_2, x_3)$) and the vertex $x_7$ (the end edge of the edge $(x_6, x_7)$) on the parent chromosome, and position all the vertices between them on the chromosome [12]. Invert as shown in Figure. 5.

| Vertex 1 of cycle | Vertex 2 of cycle | Vertex 3 of cycle | Vertex 4 of cycle | Vertex 5 of cycle | Vertex 6 of cycle |
|---|---|---|---|---|---|
| (485, 104) | (333, 149) | (397, 142) | (445, 214) | (351, 229) | (370, 181) |

*Figure 1: Schema of a chromosome.*

| (98, 319) | (255, 188) | (168, 418) | (262, 148) | (288, 72) | (337, 210) |
|---|---|---|---|---|---|

the first parent chromosome

| (255, 188) | (98, 319) | (288, 72) | (337, 210) | (168, 418) | (262, 148) |
|---|---|---|---|---|---|

the second parent chromosome

| (98, 319) | (255, 188) | (168, 418) | (288, 72) | (337, 210) | (262, 148) |
|---|---|---|---|---|---|

the child chromosome, produced from two parent chromosomes

*Figure 2: An example of a single-point crossover.*

| (98, 319) | (255, 188) | (168, 418) | (337, 210) | (288, 72) | (262, 148) |
|---|---|---|---|---|---|

the parent chromosome

| (98, 319) | (337, 210) | (168, 418) | (255, 188) | (288, 72) | (262, 148) |
|---|---|---|---|---|---|

The child chromosome, produced from the parent chromosome

*Figure 3: An example of a two-point mutation.*

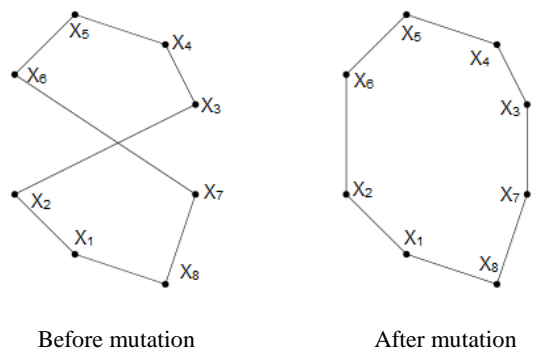

Before mutation      After mutation

*Figure 4: Removing the intersection by the second mutation operation.*

| $X_1$ | $X_2$ | $X_3$ | $X_4$ | $X_5$ | $X_6$ | $X_7$ | $X_8$ |
|---|---|---|---|---|---|---|---|

The parent chromosome

| $X_1$ | $X_2$ | $X_6$ | $X_5$ | $X_4$ | $X_3$ | $X_7$ | $X_8$ |
|---|---|---|---|---|---|---|---|

The resulting child chromosome

*Figure 5: Implementation of the second mutation on the parent choromosome.*

## 2.5 Fitness function

In the genetic algorithm, the best people are selected during different generations, and in the final generation, the best person must be selected. So we have to have a function to evaluate people's fitness. This function must be defined to meet our goal in solving the problem. The purpose of solving the problem of a straight-line embedding of a cycle is to have a planar embedding, i.e., it should not intersect the sides of a polygon and itself. For this purpose, we consider the fitness function as the sum of the number of cycle intersections with itself and the sides of a polygon. Therefore, by minimizing the fit function, the number of intersections is minimized. The fit function is shown in Formula.1.

$$F(P,Q) = C1(P) + C2(P,Q) \qquad (1)$$

$C1(P)$ represents the sum of the number of intersections P with itself and $C2(P,Q)$ represents the sum of the intersections of P with polygonal sides Q [12]. By storing the best answers in different implementations of the genetic algorithm, from all the obtained answers, the answer with the minimum number of intersections is finally selected.

## 2.6 Selection of the next generation

In the proposed genetic algorithm, we use the roulette cycle to select the next generation from the previous generation. In the roulette cycle, the best people are selected with a probability commensurate with the value of their fitness function.

## 2.7 The pseudo-code of the algorithm

By putting together all of the above, the proposed algorithm is complete. The pseudo-code of the proposed algorithm is as follows. The number of generations produced in the genetic algorithm is denoted by G. The condition for stopping the genetic algorithm at each execution is that either the number of intersections of one of the obtained cycles is zero or the number of generations produced reaches a certain value of G. In this pseudo-code, there is a main loop that, each time the loop body is executed, the genetic algorithm finds the best cycle embedding with the least intersection. The best answer of the genetic algorithm is stored in each step and at the end, the best answer is selected.

In this pseudocode in line 1, an initial population is first generated by calling the SelectFirstPopulation function. In line 2, the ComputeParentFitness function calculates the fit of each person (embedded cycle) in the initial population. Line 3 is the main loop of the genetic algorithm, which is repeated until the number of generations produced has reached a certain value of G or a cycle with zero intersection has not been produced. Inside this loop, a specific number of children are generated from the parents by calling the Reproduction function using the mutation operation and two defined mutation operations. The ComputeChildFitness function then calculates the fitness value for each child. Then, with the roulette cycle method, the best parents and children are selected as the next generation. Then, by calling the ComputeParentFitness function, the fitness level of this generation is calculated. If there is a person in this generation whose fitness is zero (number of intersections is zero) then the loop will not continue, otherwise the loop will be repeated to produce the next generation.

---

**The proposed algorithm for planar straight-line embedding of a cycle**
**Input** :simple m-polygon Q, set X of n points inside Q
**Output** :the embedding of the cycle on the points of X
1. SelectFirstPopulation
2. ComputeParentFitness
3. While G != 1000, do the followings:
   a. Reproduction
   b. ComputeChildFitness
   c. Select New Population with Roulette Wheel
   d. Current Population ← New Population
   e. ComputeParentFitness
   f. Store the BestPerson in a memory
   g. if BestPersonFitness = 0 then break,
4. EndWhile
5. Output the best resulted embedding of the Cycle stored in the memory

## 3  Results

The proposed algorithm is implemented in Python in the JetBrains PyCharm IDE. In the proposed genetic algorithm, we consider the population to be three times the number of points given inside the polygon. The number of children per generation is one third of the population. In the first version, the proposed algorithm for mutation uses only the first mutation operation, and in each reproduction, the mutation rate is 20%, the crossover rate is 80%. In the second version of the proposed algorithm, both mutation operations are used and in each reproduction, the first mutation rate is 10%, the second mutation rate is 10%, and the crossover rate is 80%. In both versions, the number of generations $G = 1000$ is considered.

We have considered parameters for measuring the behavior of algorithms, for example, to measure the effect of the number of sides on the behavior of the algorithm, we consider the problem on different polygons with different numbers of sides (ten-sides, fifteen-sides, twenty-sides and twenty-five sides) We tested. To observe the effect of the shape of a polygon on the behavior of the genetic algorithm, for each number of specified sides, we consider five different polygons with the same number of sides and average the results. Because the behavior of the algorithm is random, we run the algorithm five times for each specific fixed input instance and consider the average of the results. Therefore, for each number of specified sides, 25 tests were performed and the results were averaged. But to observe the effect of the polygon shape on the behavior of the innovative algorithm, for each given number of sides, we have considered twenty-five different polygons with the same number of sides and we have averaged on the obtained results. The number of points in the set $X$ inside the polygon is also varied (five points, ten points, fifteen points, twenty points, twenty-five points, and thirty points).

In the following, the comparison of the implementation of both versions of the genetic algorithm in terms of the average number of intersections is examined.

### 3.1  Polygons with different number of sides on fixed points

Figure. 6 shows the average number of intersections of both versions of the genetic algorithm for polygons with different numbers of sides (ten-sides, fifteen-sides, twenty-sides, and twenty-five sides) on a fixed number of points. This figure shows the output for 20 points inside each polygon. In all forms related to the results of genetic algorithm experiments, intersection 1 represents the number of intersections for the first version of the algorithm and intersection 2 represents the number of intersections for the second version of the algorithm.

As can be seen in Figure. 6, as expected, with the addition of the number of polygon sides, the average number of intersections for the first version of the algorithm has increased. This may be because the complexity of the problem becomes more difficult to find the optimal answer for the genetic algorithm as the number of sides of the polygon increases. Also, as the number of polygon sides increases, the number of planar answers in the set of answers decreases and the chances of finding them will decrease. But the interesting thing that can be seen in this figure is that the second version of the algorithm that uses the second mutation has been impressive successful and has been able to find optimal cycle embedding with zero number of intersections in all experiments.

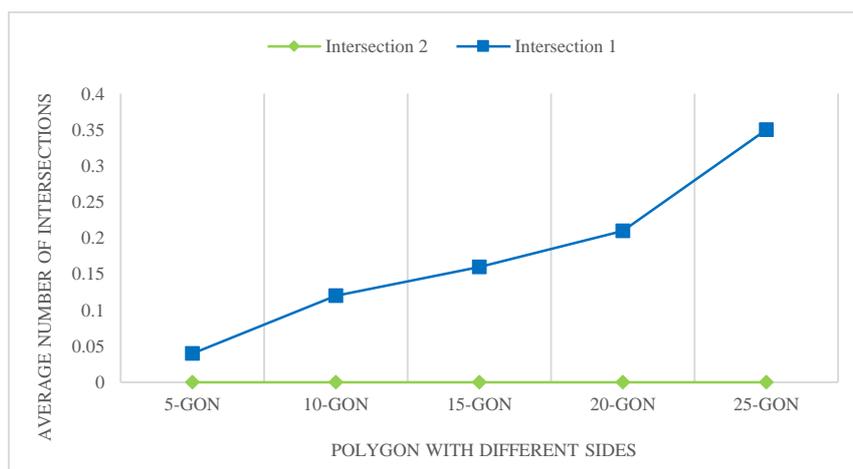

*Figure 6: The average number of intersections for polygons with different number of sides on twenty-point sets.*

### 3.2 Number of different points on fixed polygons

Figure. 7 shows the average number of intersections of both versions of the algorithm per number of different points (five points, ten points, fifteen points, twenty points, twenty-five points, and thirty points) on polygons with a fixed number of sides. These shapes show the output for a polygon.

As can be seen in Figure. 7, as the number of points increases, the average number of intersections for the first version of the algorithm increases. This may be due to the fact that as the number of points increases, the search space increases and it becomes more difficult to find the optimal answer in this space, and also as the number of points increases, the length of the cycle increases Polygons become more. But it is interesting that the second version of the algorithm has been very successful and has been able to find embedding with zero intersection in all experiments.

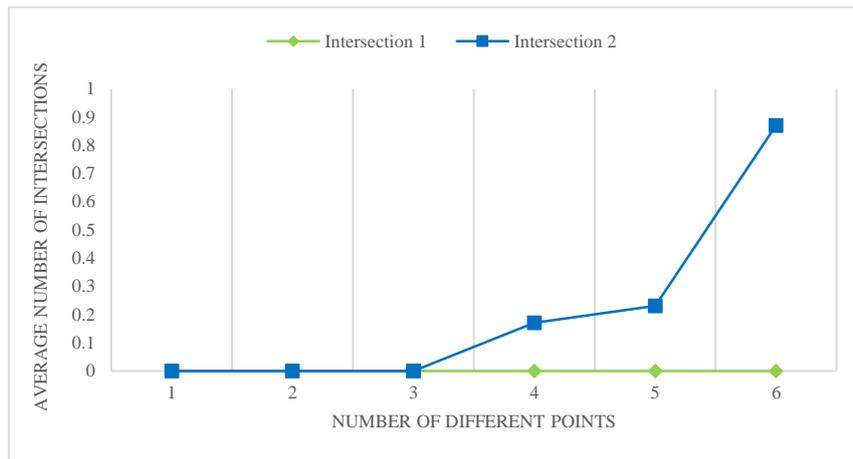

*Figure 7: The average number of intersections, for different number of points inside polygons with twenty sides.*

### 3.3 Polygons with different sides on a number of different points

Figure. 8 shows the average number of intersections of both versions of the genetic algorithm for polygons with different sides on different points. i.e., for the results shown in these figures, for each specific number of sides, the experiment is performed on polygons with the same number of sides and the number of different points, and the results are averaged.

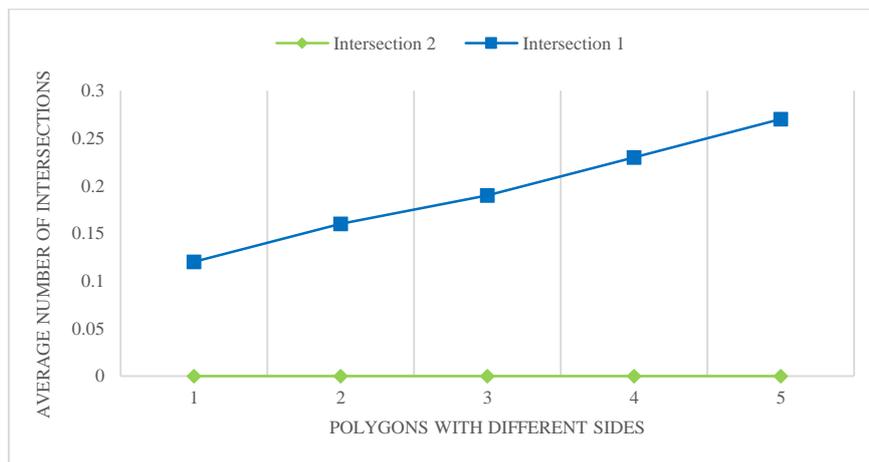

*Figure 8: The average of number of intersections, for polygons with different number of sides and different number of points.*

### 3.4 Set of points with different number of points on different polygons

Figure. 9 shows the average number of intersections of both versions of the genetic algorithm per number of different points on polygons with different numbers of sides. i.e., for the results shown in this figure, for each given number of points, the experiment is performed on several sets of points with the same number of points and different polygons with different numbers of sides, and the average is taken on the results.

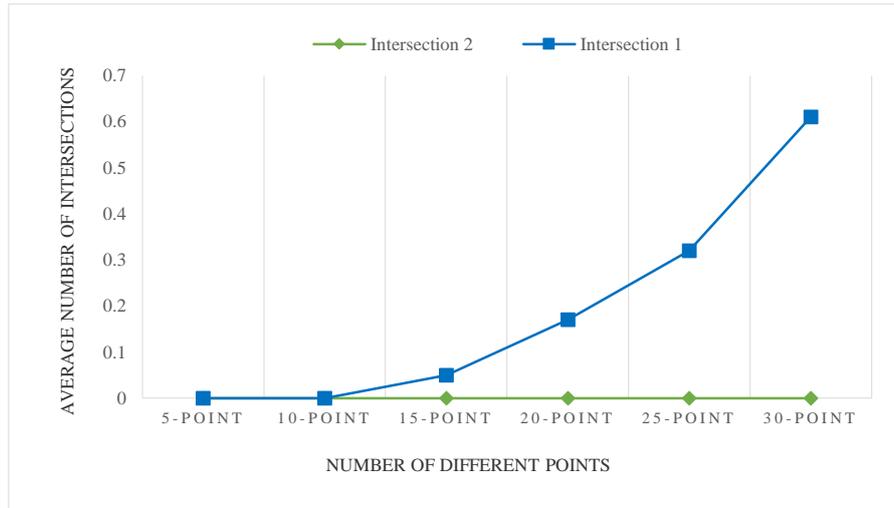

*Figure 9: The average number of intersections for some sets of points with different number of points inside different polygons.*

As can be seen in Figures. 8 and 9, similar to Figures. 6 and 7 with the same interpretation, the average number of intersections for the first version of the algorithm increases as the number of sides of the polygon increases and the points increase. But again, the number of intersections for the second version of the algorithm is zero, and this indicates the proper performance of the second mutation operation. The following are samples of the results of the implementation of the proposed genetic algorithm.

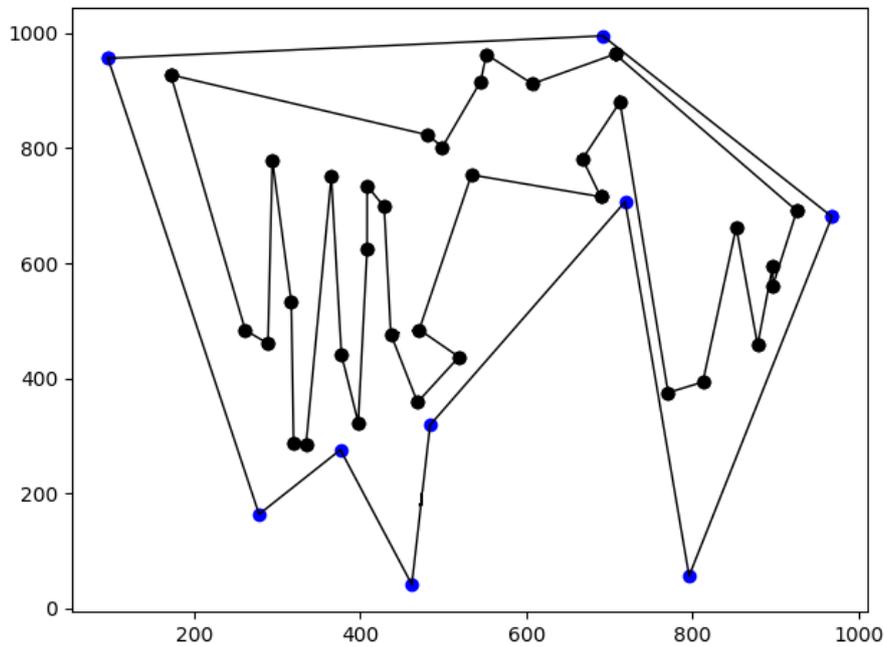

*Figure 10: Embedded cycle in 9-gon with 34 points*

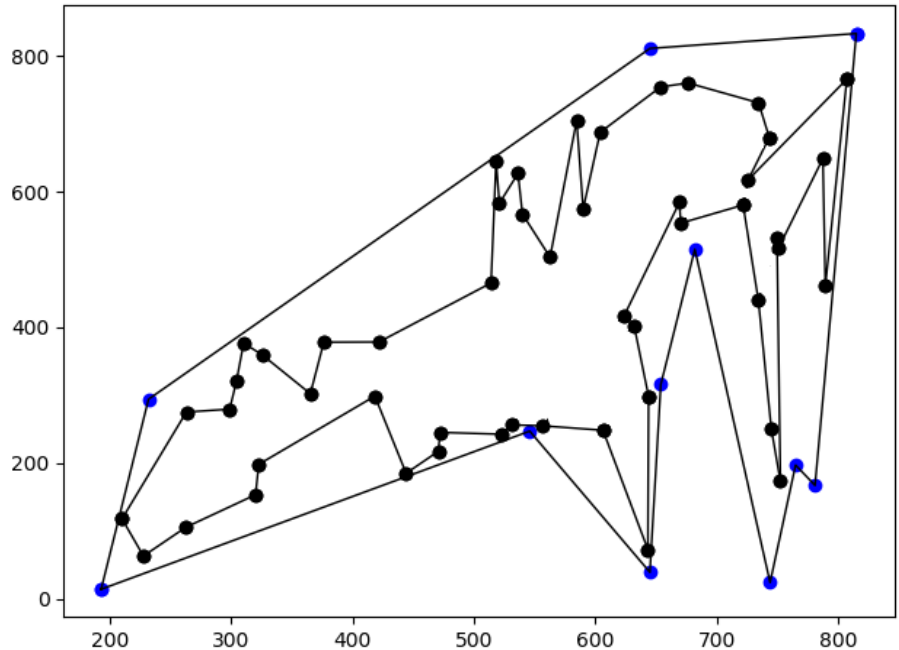

*Figure 11: Embedded cycle in 11-gon with 50 points*

## 4   Conclusions

In this paper, we present a meta-heuristic algorithm based on a genetic algorithm for straight-line embedding of a cycle with the minimum number of intersections on a point set inside simple random polygons. The work presented in this paper is a specific case of the earlier work, in which a simple path with specified length $k$ ($k \leq n$) from vertex $s$ to vertex $t$ with the least number of bends and intersections must be embedded on a set of points inside a simple polygon. The experimental results showed that the second version of the proposed genetic algorithm, which uses the second mutation operation, was much more successful than the first version, which uses a usual two-point mutation, and could reduce the number of embedding intersections to zero. This success is due to the second mutation operation, which is defined according to the objectives of the problem.